\documentclass[aps,prl,twocolumn,superscriptaddress]{revtex4-2}
\usepackage{units}
\usepackage{amsmath}
\usepackage{amssymb,amsfonts}
\usepackage{graphicx}
\usepackage{bm}
\usepackage{multirow,color,relsize,microtype}
\usepackage{dcolumn}
\usepackage{xcolor}
\usepackage{lmodern}

\newcommand{\be}{\begin{equation}}
\newcommand{\ee}{\end{equation}}

\newcommand{\im}[1]{\text{Im}\left[#1\right]}
\makeatletter
\renewcommand*\env@matrix[1][*\c@MaxMatrixCols c]{%
  \hskip -\arraycolsep
  \let\@ifnextchar\new@ifnextchar
  \array{#1}}
\makeatother

\begin{document}

\title{Robust zero modes in non-Hermitian systems without global symmetries}

\author{Jose D. H. Rivero}
\affiliation{College of Staten Island, CUNY, Staten Island, New York 10314, USA}
\affiliation{The Graduate Center, CUNY, New York, NY 10016, USA}

\author{Courtney Fleming}
\affiliation{College of Staten Island, CUNY, Staten Island, New York 10314, USA}
\affiliation{The Graduate Center, CUNY, New York, NY 10016, USA}

\author{Bingkun Qi}
\affiliation{College of Staten Island, CUNY, Staten Island, New York 10314, USA}
\affiliation{The Graduate Center, CUNY, New York, NY 10016, USA}

\author{Liang Feng}
\affiliation{Department of Materials Science and Engineering, University of Pennsylvania, Philadelphia, PA 19104, USA}

\author{Li Ge}
\affiliation{College of Staten Island, CUNY, Staten Island, New York 10314, USA}
\affiliation{The Graduate Center, CUNY, New York, NY 10016, USA}
\email{li.ge@csi.cuny.edu}
\date{\today}

\begin{abstract}
We present an approach to achieve zero modes in lattice models that do not rely on any symmetry or topology of the bulk, which are robust against disorder in the bulk of \textit{any} type and strength. Such symmetry-free zero modes (SFZMs) are formed by attaching a single site or small cluster with zero mode(s) to the bulk, which serves as the ``nucleus'' that expands to the entire lattice. We identify the requirements on the couplings between this boundary and the bulk, which reveals that this approach is intrinsically non-Hermitian. We then provide several examples with either an arbitrary or structured bulk, forming spectrally embedded zero modes in the bulk continuum, midgap zero modes, and even restoring the ``zeroness'' of coupling or disorder-shifted topological corner states. Focusing on viable realizations using photonic lattices, we show that the resulting SFZM can be observed as the single lasing mode when optical gain is applied to the boundary. 
\end{abstract}

\maketitle

Zero modes occupy an indisputable place of interest in physics at many different levels: They appear as localized boundary states in topological insulators \cite{hasan_colloquium:_2010,Qi2011}, Majorana fermions in superconductors and semiconductors \cite{alicea_new_2012,beenakker_random-matrix_2015,Elliott2015}, corner states in higher-dimensional multipole insulators \cite{Benalcazar2017,Peterson2018,ElHassan2019}, single-mode lasers with narrow spectral lines \cite{Ge2017,Zhao2018,hodaei2014parity,poli_selective_2015,peng_parity-time-symmetric_2014}, and localized defect modes in photonic crystals \cite{mertens2005tunable,kuzmiak1998localized,painter1999defect,villeneuve1996microcavities,pan_photonic_2018}. Many advantages come with the ability to excite these states in the middle of band gaps, and the symmetry or topological properties of the underlying system can be judiciously implemented to protect those states.

It is widely accepted that symmetries play a special role in the construction of these zero modes in complex systems, as they introduce a privileged energy (i.e., ``zero energy'') that serves as a reference for all the energies of the system. In condensed matter systems, this zero energy is often the Fermi energy \cite{kittel1996introduction}, and in coupled photonic systems, the resonant frequency of interest in a single cavity or waveguide. Two well-known symmetries that warrant zero modes in Hermitian systems are chiral and particle-hole symmetries, which lead to a symmetric energy spectrum about zero \cite{chaikin1995principles}. The study of non-Hermitian physics has also exploited and extended these two symmetries in the presence of dissipation, where gain and loss can be engineered to produce zero modes with unique properties \cite{Feng2017}. As the spectrum becomes complex, chiral symmetry \cite{rivero2021chiral,yoshida2019symmetry,lee2016anomalous,yin2018geometrical,Kawabata2019}, as well as pseudo-chirality \cite{rivero_pseudochirality_2020}, still hosts paired modes with energy $E$ and $-E$; particle-hole symmetry \cite{Ge2017,qi2018defect,kawabata2019topological,okugawa2019topological,wu2019symmetry}, along side anti-PT symmetry \cite{ge_antisymmetric_2013,zhang_synthetic_2020} and pseudo-anti-Hermiticity \cite{scolarici_pseudoanti-hermitian_2002}, now produces mode pairs with energy $E$ and $-E^*$, where the asterisk denotes the complex conjugation as usual.

As these symmetries are \text{global} properties of the underlying system, the zero modes they induce are easily lifted in the presence of \text{local} perturbations that do not preserve these symmetries. For example, the zero modes of a chiral-symmetric lattice disappear in the presence of strong onsite disorder. While a lattice with non-Hermitian particle-hole symmetry has enhanced protection in comparison from gain and loss disorder \cite{Ge2017}, its zero modes are still eliminated with onsite frequency detunings. Although one may mitigate the problem by employing topological protection \cite{Chiu2016,Malzard2015,schomerus2013topologically,weimann2017topologically}, the resulting modes typically still have energy shifted away from zero, causing uncertainties in the identification and characterization of the zero modes.

Meanwhile, a zero mode in a single site or a small cluster is relatively easy to form, since the zero energy, for example in a photonic lattice, is chosen as that of a resonant frequency in a single cavity as previously mentioned. Therefore, can one utilize such a zero mode as the ``nucleus'' and expand it to the entire lattice?

In this work, we elucidate this vastly different approach to realize a robust zero mode in lattice models, which does not depend on the symmetry or topological properties of the bulk. By attaching the aforementioned nucleus to an arbitrary bulk and referring to the former as the boundary, we identify the conditions needed on the bulk-boundary couplings to form such a symmetry-free zero mode (SFZM). 
It is worth noting that non-Hermiticity is intrinsic in our approach, either in the form of asymmetric couplings \cite{Longhi_robust_2015,zhang_observation_2021,wang_non-hermitian_2022,franca_non-hermitian_2022,gao_two-dimensional_2023} or a non-Hermitian boundary.

A distinct property of SFZMs is their robustness against \textit{any} type of disorder in the bulk Hamiltonian. Furthermore, they can exist in systems both with and without a bulk energy gap at zero energy, which enables us to \textit{freely} form spectrally embedded zero modes similar to bound states in the continuum (BIC) of the bulk \cite{hsu_bound_2016,longhi_bound_2014}, midgap zero modes, and even restore the ``zeroness'' of coupling or disorder-shifted topological corner states. Therefore, while the mechanism we introduce to achieve a zero mode in the bulk does not rely on global symmetries, it can be applied to systems with global symmetries and topological protection as well. In this case, our approach provides additional robustness to the underlying system, stabilizing its energy against symmetry-breaking perturbations. Focusing on viable realizations using photonic lattices, we also show that an SFZM can be observed as the single lasing mode when optical gain is applied to the boundary. Such an approach can benefit many applications, especially those in optics and photonics that rely on a source of a fixed frequency for spectroscopy, metrology, and sensing.

\begin{figure}[t]
	\centering \includegraphics[width = \linewidth]{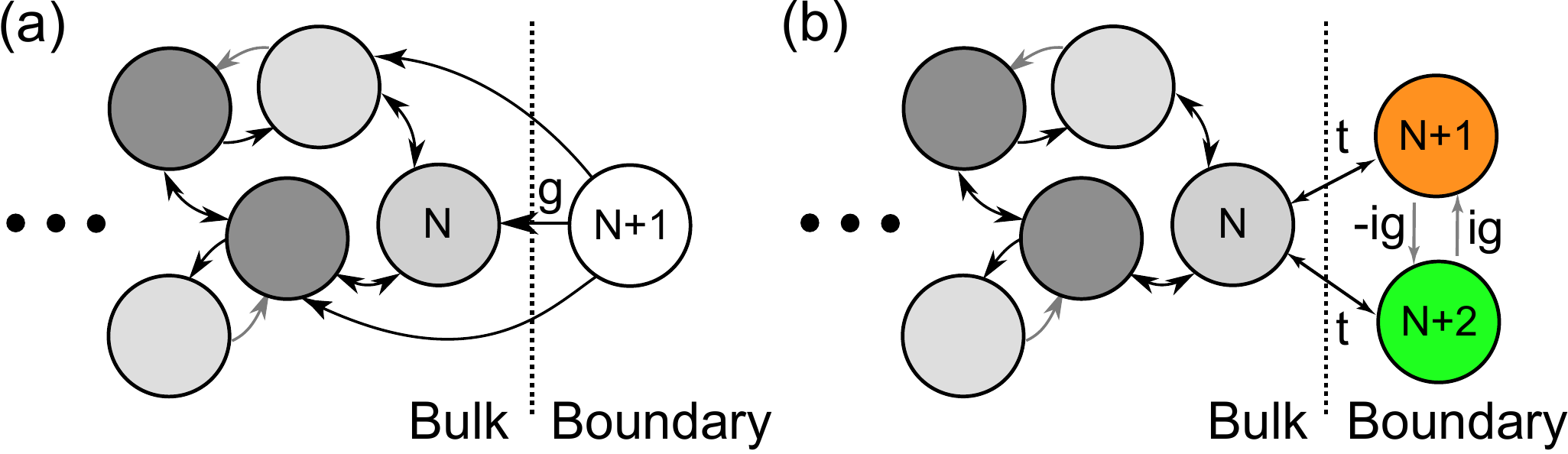}
	\caption{Schematics of systems hosting an SFZM, where the boundary is a single cavity in (a) and a small cluster in (b). 
	}\label{fig:generic}
\end{figure}

Below we consider the following Hamiltonian
\begin{equation}
	H = \begin{pmatrix}
		H_B & V\\
		U & H_0
	\end{pmatrix}
	\label{eq:H}
\end{equation}
consisting of the bulk ($H_B$), the boundary ($H_0$), and their couplings $U,V$. 
Our boundary serves as the nucleus and hosts zero mode(s) itself (\textbf{Condition 0}), i.e., $H_0\phi_i=0$. By partitioning the wave function of the whole system as $\Psi = \psi_B\oplus\psi_0$, a zero mode of the system satisfies
\be
H_0'\psi_0\equiv(H_0-UH_B^{-1}V)\psi_0 =0,\;\;\;\psi_B = -H_B^{-1}V\psi_0.\label{eq:partition}
\ee
Our goal is to achieve zero mode(s) inside a bulk without pre-existing ones, and hence we have taken $H_B$ to be invertible. 
We also note that the second relation in Eq.~(\ref{eq:partition}) indicates that the zero-mode wave function in the bulk (i.e., $\psi_B$) can be obtained by propagating, via the bulk Green's function $G_B(E)=(E-H_B)^{-1}$ at the zero energy, the couplings from the boundary. 


To identify a set of general conditions that enables an SFZM, we use the property that a Hamiltonian always has paired left and right eigenstates of the same energies \cite{bronson1991matrix,linearAlgebra}. The first relation in Eq.~(\ref{eq:partition}) is then satisfied if we can construct a simple left eigenstate of $H'_0$ with $E=0$, which should be independent of the bulk Hamiltonian $H_B$. This goal can be achieved by requiring $\tilde{\phi}_i^T U=0$ (\textbf{Condition 1}), where $\tilde{\phi}_i^T$ is the left eigenstate of a boundary zero mode, defined by $\tilde{\phi}_i^T H_0=0$. This condition can be easily verified by noting $\tilde{\phi}_i^T(H_0-UH_B^{-1}V)=0$. In other words, $\tilde{\phi}_i^T$ and $\psi_0$ are the left and right eigenstates of $H_0'$ with $E=0$. In contrast, we note that the boundary zero mode $\phi_i$ and the SFZM inside the boundary (i.e.,$\psi_0$) are not simply related in general.

The resulting zero mode of $H$, however, has a vanished amplitude in the bulk if $V \phi_i=0$: $V \phi_i=0$ indicates that $H_0'\phi_i=0$, i.e., the right eigenstate of $H_0'$ with $E=0$, denoted by $\psi_0$ in Eq.~(\ref{eq:partition}), is simply given by $\phi_i$. Using $V \psi_0=V \phi_i=0$ in Eq.~(\ref{eq:partition}), we then find $\psi_B=0$. Therefore, to achieve an SFZM that extends to the bulk, we require $V \phi_i \neq 0$ (\textbf{Condition 2}). An alternative derivation of Conditions 1 and 2 is given in Ref.~\cite{SM} using the Jordan normal form of $H_0$, where the case with a pre-existing bulk zero mode is also discussed.

Conditions 1 and 2 imply that the system that hosts an SFZM is \textit{intrinsically} non-Hermitian: If the opposite were true, then $H_0$ must be Hermitian as well, leading to $\tilde{\phi}_i^T=\phi_i^\dagger$. At the same time, we have $U=V^\dagger$, and by taking the Hermitian conjugate of Condition 1, we find $V\phi_i=0$ that violates Condition 2. Similarly, Conditions 1 and 2 also imply that $H$ is asymmetric, meaning that the left and right eigenstates of the SFZM are distinct. This observation provides an intuitive understanding of the SFZM: By imposing Condition 1, we achieve not just a left zero mode of $H_0'$ but also that of $H$, with the simple wave function $\tilde{\Psi}^T=0\oplus\tilde{\phi}_i^T$ that vanishes in the bulk and satisfies $\tilde{\Psi}^TH=0$. The corresponding and distinct right eigenstate of $H$ is then the SFZM, which extends to the bulk under Condition 2.

As an example, we first consider the simplest case where the proposed SFZM appears [Fig.~\ref{fig:generic}(a)]. It features a single boundary site with $H_0=0$, satisfying Condition 0 with a boundary zero mode given by $\tilde{\phi}_i={\phi}_i=1$. Then Condition 1 dictates that the coupling $U$ must be a zero row vector, i.e., no couplings from the bulk to the boundary. The existence of a zero mode in the whole system is then apparent: 
the bulk cannot act back on the boundary, and hence the frequency of the latter's zero mode is unchanged.

\begin{figure}[t]
	\centering \includegraphics[width = \linewidth]{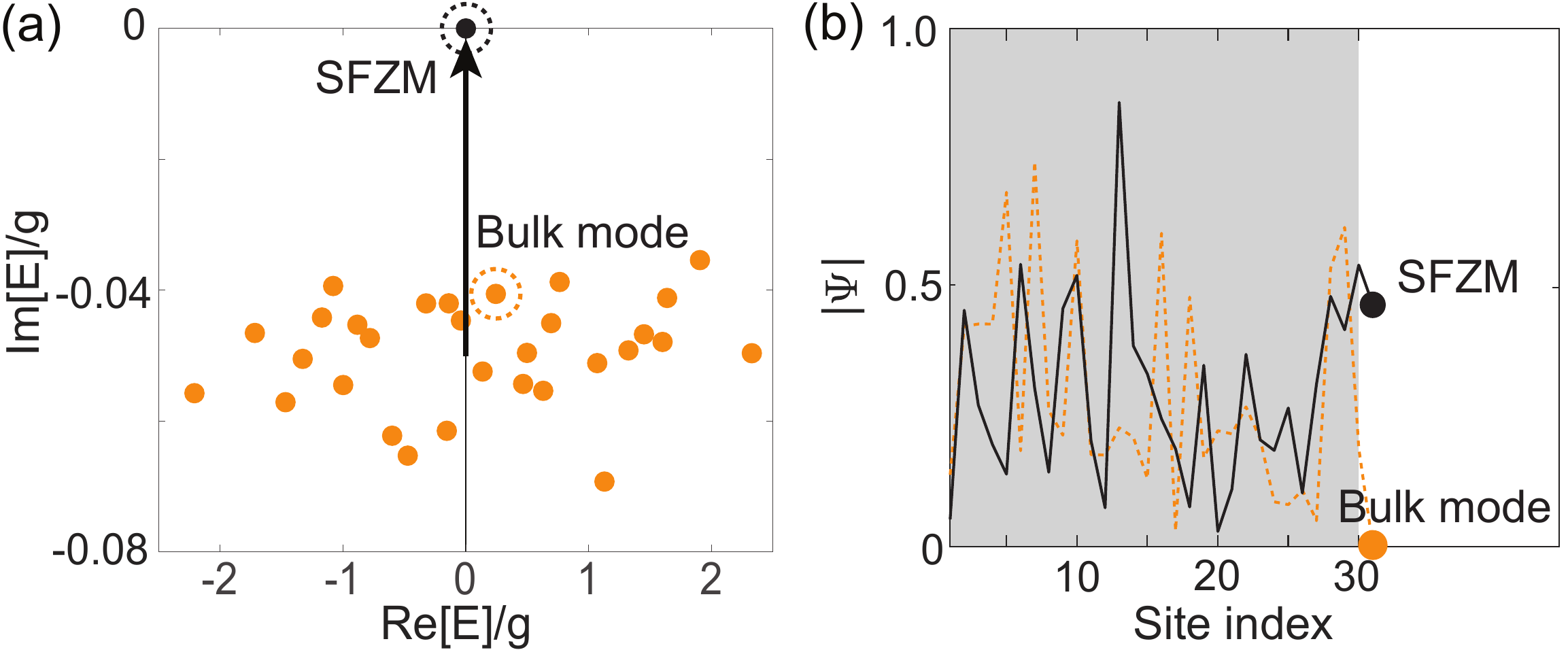}
	\caption{(a) Representative complex energy spectrum for the configuration shown in Fig.~\ref{fig:generic}(a) with 30 bulk lattice sites. Both onsite energy and couplings in the bulk are uniformly distributed in $[-g/2,g/2]$, 
	whereas onsite loss is uniformly distributed in $[-g/10,0]$. 
	Black and orange dots mark the SFZM and bulk nonzero modes. Black arrow shows the trajectory of the SFZM when the cavity loss in the boundary is gradually compensated by an increasing pump. (b) Spatial profiles of the SFZM and the circled bulk mode in (a). Shaded area shows the bulk.}\label{fig:singleB}
\end{figure}

To ensure that the resulting SFZM extends into the bulk, Condition 2 requires that the coupling $V$ is not a zero vector, which would otherwise isolate the boundary from the bulk. 
We also note that the required unidirectional couplings from the boundary to the bulk can be realized, for example, using spiral waveguides evanescently coupled to micro-ring resonators \cite{gao_two-dimensional_2023}. For easy implementation, coupling the boundary cavity to one in the bulk [e.g., the $N$th cavity in Fig.~\ref{fig:generic}(a)] 
 is sufficient to warrant the SFZM. Importantly, the bulk Hamiltonian $H_B$ can be completely arbitrary and does not affect the existence of the SFZM [Fig.~\ref{fig:singleB}(a)].


We note that this SFZM is the \textit{only} eigenstate of the whole system that has a finite amplitude at the boundary site. The other eigenstates are simply given by all the $N$ eigenstates of the bulk, with a zero amplitude at the boundary [see, for example, the dashed line in Fig.~\ref{fig:singleB}(b)], due to the prohibited couplings from the bulk to the boundary. These properties facilitate the observation of the SFZM in an active device (e.g., a laser \cite{Feng2017}), which exhibits mode(s) with the smallest loss or highest gain. Here by including sufficient loss in the bulk, whether uniform or distributed, all the non-zero modes in the whole system are in the lower half of the complex energy plane [Fig.~\ref{fig:singleB}(a)]. As a result, only the SFZM is on the real axis and similar spectrally to a BIC, especially as the density of states increases with the bulk size. 

This picture, of course, is simplified because we have not included the cavity loss in the boundary cavity itself. However, even when we consider a decay rate in this cavity comparable to the bulk modes ($\kappa/g=-0.05i$), the SFZM can be observed directly as the only lasing mode in this system. This behavior is shown by the black arrow in Fig.~\ref{fig:singleB}(a): We pump only this cavity that reduces its effective cavity decay, i.e., $\kappa\rightarrow\kappa-\gamma$. When the gain $\gamma$ compensates this cavity loss, i.e., $\gamma=\kappa$, the SFZM reach its lasing threshold, i.e., $\im{E}=0$, while the bulk modes \textit{do not} move upward at all towards their respective lasing thresholds in this process, since their wave functions do not extend into the boundary cavity where the gain is applied.



\begin{figure}[t]
	\centering \includegraphics[width = \linewidth]{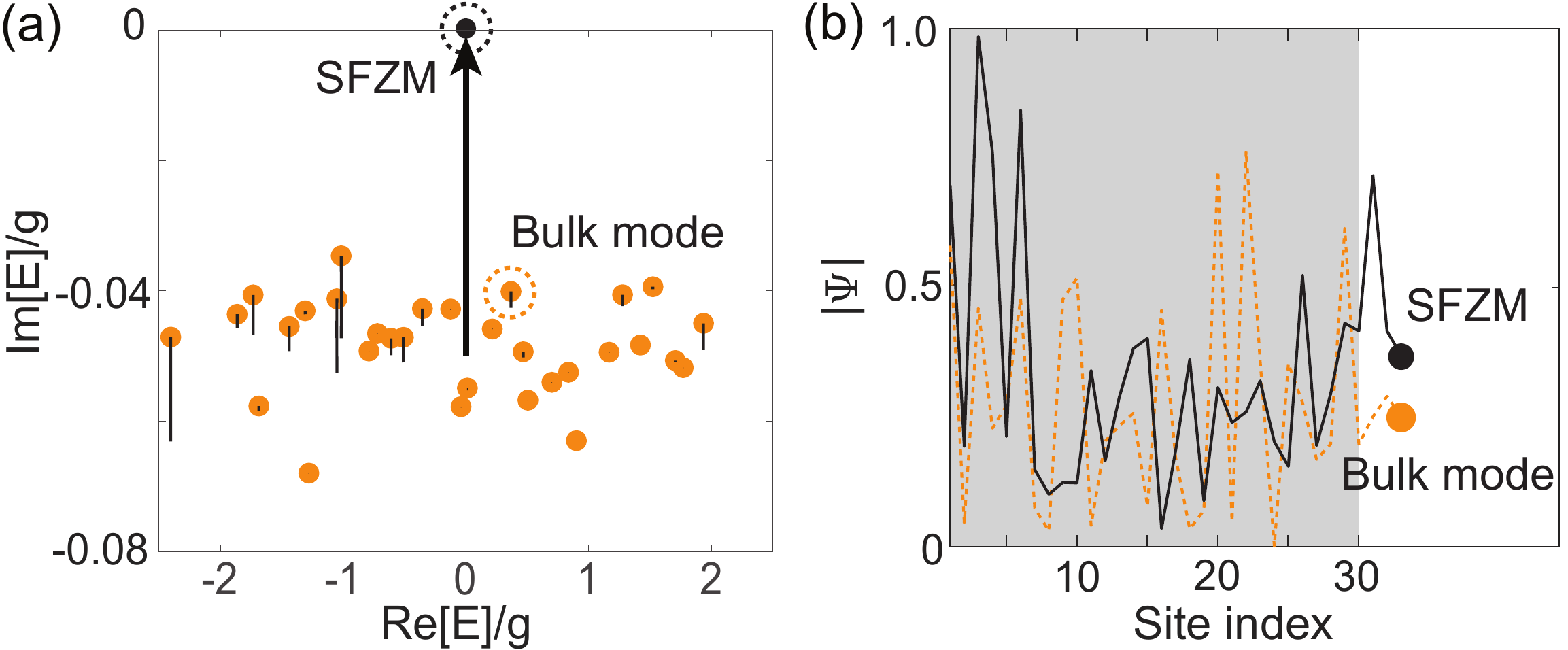}
	\caption{Same as Fig.~\ref{fig:singleB} but with a three-site boundary and another arbitrary bulk. $k=g$, $k'=-g$, and $q=2g$ are used.}\label{fig:3B}
\end{figure}

As another example, consider a three-cavity boundary
\be
	H_0 = \begin{pmatrix}
		0 & g & 0\\
		g & 0 & g\\
		0 & g & 0
	\end{pmatrix}, \,
	U = \begin{pmatrix}
		\cdots & 0 & k\\
		\cdots & 0 & q\\
		\cdots & 0 & k
	\end{pmatrix},\,
V = \begin{pmatrix}
		\vdots & \vdots & \vdots \\
		0 & 0 & 0 \\
		k' & q & k
	\end{pmatrix}.\nonumber
\ee
The boundary zero mode is given by $\tilde{\phi}_i={\phi}_i=[1,0,-1]^T$, and we have required asymmetric couplings $k'\neq k$ to meet Condition 2 (instead of unidirectional couplings as in the previous example). 
In this case, the SFZM is no longer the only state with finite amplitudes in the boundary [see Fig.~\ref{fig:3B}(b)], but we can still achieve a BIC-like spectrum [see Fig.~\ref{fig:3B}(a)] by including sufficient loss in the bulk. This SFZM can then again be observed as the single lasing mode when only the optical cavities in the boundary are pumped to compensate for their losses, which makes the imaginary parts of the diagonal elements in $H_0$ back to zero. Some of the bulk nonzero modes, due to their finite amplitudes in the boundary cavities, experience reduced losses in this process [see the short vertical black lines in Fig.~\ref{fig:3B}(a)], but they are still far away from their lasing thresholds (i.e., the real axis) \cite{SM}.

\begin{figure}[b]
	\centering \includegraphics[width = \linewidth]{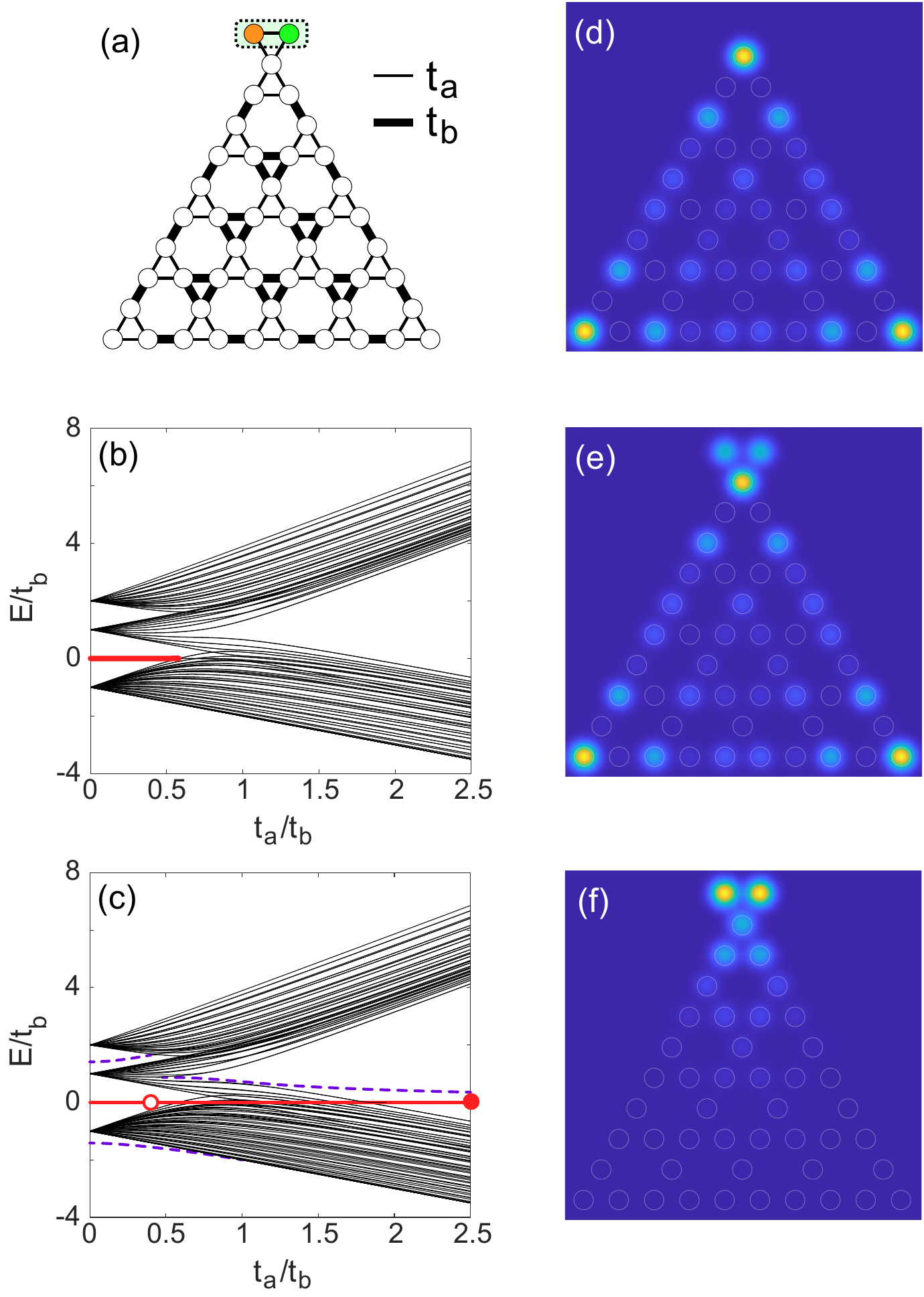}
	\caption{(a) Schematics of a breathing Kagome lattice, with the non-Hermitian boundary attached. (b,c) Band diagram without and with the boundary. $g=t=t_b$ and 20 rows of cavities are used. (d) Symmetric higher-order corner state in the topological gap at $t_a/t_b=0.4$ without the boundary. (e,f) The wave functions of the SFZMs at $t_a/t_b=0.4,2.5$ [open and closed dots in (c)]. 10 rows of cavities are used in (d-f) for a compact illustration.}\label{fig:Kagome}
\end{figure}

In both examples above, we have used non-Hermitian couplings to couple an effectively Hermitian boundary to the bulk. Below we present another example where the bulk-boundary couplings are Hermitian but the boundary itself is non-Hermitian instead. Consider the schematic shown in Fig.~\ref{fig:generic}(b), where the boundary consists of two cavities at zero frequency. 
These two cavities have opposite gain and loss (i.e., $\pm ig$) and are connected by couplings of the same strength ($\pm ig$). 
Together with the last cavity in the bulk, we denote the Hamiltonian of this subsystem by
\be
H_d =
\left(
\begin{array}{c|c c}
\omega_N & t & t \\ \hline
t & -ig & ig \\
t & -ig & ig
\end{array}
\right),\label{eq:1}
\ee
where $\omega_N$ is the mode frequency in the last bulk cavity ($N$) and $t$ is its coupling to the two boundary cavities. Since the latter are not coupled to other cavities in the bulk, the rest of $U$ and $V$ are all zero. Conditions 0--2 are again satisfied, with the boundary zero mode given by the left and right eigenstates $\tilde{\phi}_i^T=[1, -1]$, ${\phi}_i=[1, 1]^T$. Clearly, $H_d$ (and $H$) has a vanished determinant due to the repeated rows, and in turn, also an SFZM. We note that although in this case the boundary zero mode is at the exceptional point (EP) \cite{miri_exceptional_2019} of $H_0$, this condition is not required by Hermitian boundary and bulk couplings \cite{SM}. Furthermore, even when the boundary zero mode is at an EP, the formed SFZM is not an EP in general, as can be easily checked using $H_d$ given above.

Below we exemplify SFZMs in this configuration using a triangular ``breathing'' Kagome lattice \cite{Kagome1,Kagome2}, which has alternate nearest-neighbor couplings $t_a,t_b$ [see Fig.~\ref{fig:Kagome}(a)] that enable topologically protected higher-order corner states. Due to the lack of chiral symmetry and the finite length of their tails, they couple and have energies slightly off $E=0$, represented by the thick red line in the band diagram when $t_a/t_b\lesssim0.5$ [Fig.~\ref{fig:Kagome}(b)]. A trivial bandgap also opens for $t_a/t_b\gtrsim1$. 

With the non-Hermitian boundary attached, it is expected that some defect modes will appear, and they are shown as the dashed lines in Fig.~\ref{fig:Kagome}(c). Most noticeably though, now there is a zero mode independent of the value of $t_a$ (thin red line), whether it is in the topological gap on the left [Fig.~\ref{fig:Kagome}(e)], through the bulk band in the middle, or in the trivial gap on the right [Fig.~\ref{fig:Kagome}(f)]. In addition, unlike the original midgap states whose energies are shifted from zero (e.g., $E=0.01t_b$ for the symmetric one in Fig.~\ref{fig:Kagome}(d) at $t_a/t_b=0.4$), this SFZM is exactly at $E=0$. In other words, this process of boundary nucleation in fact restores the ``zeroness'' of one topological corner state, with (essentially) the same wave function in the bulk \cite{bibnote}. 


\begin{figure}[t]
	\centering \includegraphics[width = \linewidth]{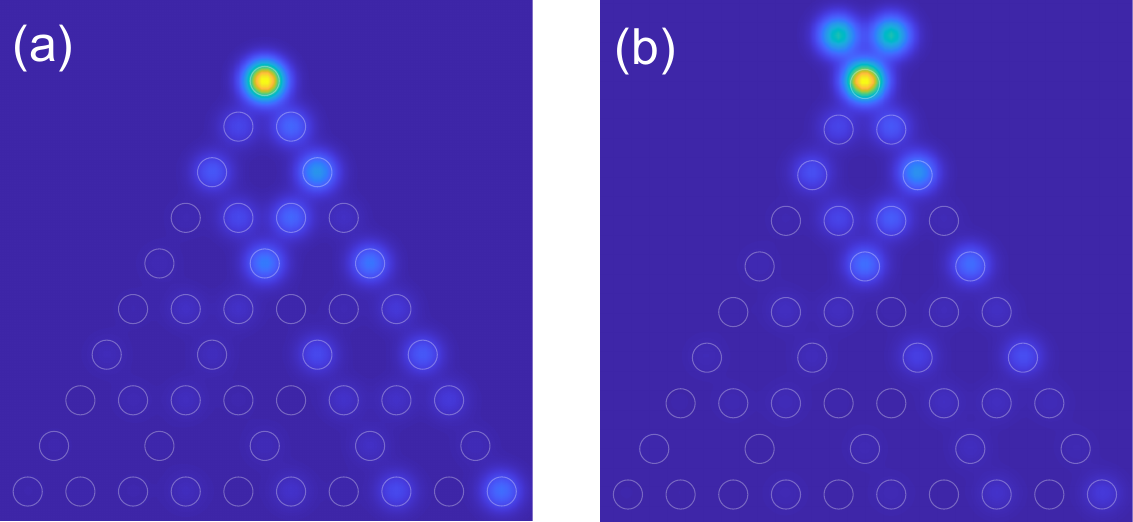}
	\caption{(a) Wave function of a near-zero mode without the boundary and with disorder. Onsite and coupling disorders are uniform in the ranges $[-0.1,0.1]t_b$ and $[-0.4,0.4]t_b$, respectively. (b) Corresponding SFZM with the boundary at the top.}\label{fig:Kagome_disordered}
\end{figure}

To further stress that the energy of the SFZM is robust against any disorder in the bulk, we now introduce both onsite and coupling disorders to the bulk and focus on the configuration with $t_a/t_b=0.4$. Before we attach the boundary, the mode closest to $E=0$ in one disorder realization is shown in Fig.~\ref{fig:Kagome_disordered}(a). It resembles the corner state shown in Fig.~\ref{fig:Kagome}(d) without disorder because of its topological protection, with the highest amplitude now at just one corner due to the break of the $C_{3v}$ symmetry. Its energy $E/t_b=-0.0365$ is negative, partly due to mixing with a hexagon plaquette right beneath the apex, which corresponds to one localized compact state in the flat band at $E=-(t_a+t_b)$ in the absence of disorder. Despite these changes to its wave function, the attached boundary again restores the zeroness of this disorder-shifted state, while maintaining its spatial profile in the bulk [Fig.~\ref{fig:Kagome_disordered}(b)]. The latter occurs because this disordered state is mainly propagated by the bulk Green's function [see Eq.~(\ref{eq:partition})] from the top corner, with or without the boundary cluster, and the energy difference in these two cases is small.

In summary, we have introduced an approach to realize zero modes in lattice models that do not rely on any symmetry or topology of the bulk. They originate from a boundary with zero mode(s) that serve as the nucleus of these SFZMs, and they are robust against disorder of any kind and strength in the bulk Hamiltonian. While certain perturbations in the bulk can also change bulk-boundary couplings \cite{SM} beyond the lattice model, the system is much less error-prone with the addition of the proposed boundary cluster, given the potentially vast number of lattice sites in the bulk versus the small numbers of boundary sites and bulk-boundary couplings we utilize.

We have exemplified these SFZMs using both arbitrary and structured bulks. Though not shown in the latter, these SFZMs can be observed as the single lasing mode on an active photonic lattice. We also note that if there are $n$ boundary zero modes with distinct wave functions, then each can serve as a nucleus and form an SFZM, leading to a maximum of $n$ SFZMs simultaneously. We have also excluded bulks with pre-existing zero mode(s) because they may leave the system without an SFZM \cite{SM}, unlike the near-zero modes in Figs.~\ref{fig:Kagome} and \ref{fig:Kagome_disordered}. Our finding provides a different perspective on zero modes in both mundane and topological systems, which can also be used to restore the zero energy of coupling or disorder-shifted topological states.  

\acknowledgments
This project is supported by National Science Foundation under Grants No. PHY-1847240 and No. ECCS-1846766.

\bibliography{references}

\begin{thebibliography}{54}%
\makeatletter
\providecommand \@ifxundefined [1]{%
 \@ifx{#1\undefined}
}%
\providecommand \@ifnum [1]{%
 \ifnum #1\expandafter \@firstoftwo
 \else \expandafter \@secondoftwo
 \fi
}%
\providecommand \@ifx [1]{%
 \ifx #1\expandafter \@firstoftwo
 \else \expandafter \@secondoftwo
 \fi
}%
\providecommand \natexlab [1]{#1}%
\providecommand \enquote  [1]{``#1''}%
\providecommand \bibnamefont  [1]{#1}%
\providecommand \bibfnamefont [1]{#1}%
\providecommand \citenamefont [1]{#1}%
\providecommand \href@noop [0]{\@secondoftwo}%
\providecommand \href [0]{\begingroup \@sanitize@url \@href}%
\providecommand \@href[1]{\@@startlink{#1}\@@href}%
\providecommand \@@href[1]{\endgroup#1\@@endlink}%
\providecommand \@sanitize@url [0]{\catcode `\\12\catcode `\$12\catcode
  `\&12\catcode `\#12\catcode `\^12\catcode `\_12\catcode `\%12\relax}%
\providecommand \@@startlink[1]{}%
\providecommand \@@endlink[0]{}%
\providecommand \url  [0]{\begingroup\@sanitize@url \@url }%
\providecommand \@url [1]{\endgroup\@href {#1}{\urlprefix }}%
\providecommand \urlprefix  [0]{URL }%
\providecommand \Eprint [0]{\href }%
\providecommand \doibase [0]{https://doi.org/}%
\providecommand \selectlanguage [0]{\@gobble}%
\providecommand \bibinfo  [0]{\@secondoftwo}%
\providecommand \bibfield  [0]{\@secondoftwo}%
\providecommand \translation [1]{[#1]}%
\providecommand \BibitemOpen [0]{}%
\providecommand \bibitemStop [0]{}%
\providecommand \bibitemNoStop [0]{.\EOS\space}%
\providecommand \EOS [0]{\spacefactor3000\relax}%
\providecommand \BibitemShut  [1]{\csname bibitem#1\endcsname}%
\let\auto@bib@innerbib\@empty
\bibitem [{\citenamefont {Hasan}\ and\ \citenamefont
  {Kane}(2010)}]{hasan_colloquium:_2010}%
  \BibitemOpen
  \bibfield  {author} {\bibinfo {author} {\bibfnamefont {M.~Z.}\ \bibnamefont
  {Hasan}}\ and\ \bibinfo {author} {\bibfnamefont {C.~L.}\ \bibnamefont
  {Kane}},\ }\bibfield  {title} {\bibinfo {title} {Colloquium: {Topological}
  insulators},\ }\href {https://doi.org/10.1103/RevModPhys.82.3045} {\bibfield
  {journal} {\bibinfo  {journal} {Rev. Mod. Phys.}\ }\textbf {\bibinfo {volume}
  {82}},\ \bibinfo {pages} {3045} (\bibinfo {year} {2010})}\BibitemShut
  {NoStop}%
\bibitem [{\citenamefont {Qi}\ and\ \citenamefont {Zhang}(2011)}]{Qi2011}%
  \BibitemOpen
  \bibfield  {author} {\bibinfo {author} {\bibfnamefont {X.~L.}\ \bibnamefont
  {Qi}}\ and\ \bibinfo {author} {\bibfnamefont {S.~C.}\ \bibnamefont {Zhang}},\
  }\bibfield  {title} {\bibinfo {title} {Topological insulators and
  superconductors},\ }\href@noop {} {\bibfield  {journal} {\bibinfo  {journal}
  {Rev. Mod. Phys.}\ } \textbf {\bibinfo {volume}{83}},\ \bibinfo {pages} {1057} (\bibinfo {year} {2011})}\BibitemShut {NoStop}%
\bibitem [{\citenamefont {Alicea}(2012)}]{alicea_new_2012}%
  \BibitemOpen
  \bibfield  {author} {\bibinfo {author} {\bibfnamefont {J.}~\bibnamefont
  {Alicea}},\ }\bibfield  {title} {\bibinfo {title} {New directions in the
  pursuit of {Majorana} fermions in solid state systems},\ }\href
  {https://doi.org/10.1088/0034-4885/75/7/076501} {\bibfield  {journal}
  {\bibinfo  {journal} {Rep. Prog. Phys.}\ }\textbf {\bibinfo {volume} {75}},\
  \bibinfo {pages} {076501} (\bibinfo {year} {2012})}\BibitemShut {NoStop}%
\bibitem [{\citenamefont {Beenakker}(2015)}]{beenakker_random-matrix_2015}%
  \BibitemOpen
  \bibfield  {author} {\bibinfo {author} {\bibfnamefont {C.}~\bibnamefont
  {Beenakker}},\ }\bibfield  {title} {\bibinfo {title} {Random-matrix theory of
  {Majorana} fermions and topological superconductors},\ }\href
  {https://doi.org/10.1103/RevModPhys.87.1037} {\bibfield  {journal} {\bibinfo
  {journal} {Rev. Mod. Phys.}\ }\textbf {\bibinfo {volume} {87}},\ \bibinfo
  {pages} {1037} (\bibinfo {year} {2015})}\BibitemShut {NoStop}%
\bibitem [{\citenamefont {Elliott}\ and\ \citenamefont
  {Franz}(2015)}]{Elliott2015}%
  \BibitemOpen
  \bibfield  {author} {\bibinfo {author} {\bibfnamefont {S.~R.}\ \bibnamefont
  {Elliott}}\ and\ \bibinfo {author} {\bibfnamefont {M.}~\bibnamefont
  {Franz}},\ }\bibfield  {title} {\bibinfo {title} {{C}olloquium: {M}ajorana
  fermions in nuclear, particle, and solid-state physics},\ }\href
  {https://doi.org/10.1103/RevModPhys.87.137} {\bibfield  {journal} {\bibinfo
  {journal} {Rev. Mod. Phys.}\ }\textbf {\bibinfo {volume} {87}},\ \bibinfo
  {pages} {137} (\bibinfo {year} {2015})}\BibitemShut {NoStop}%
\bibitem [{\citenamefont {Benalcazar}\ \emph {et~al.}(2017)\citenamefont
  {Benalcazar}, \citenamefont {Bernevig},\ and\ \citenamefont
  {Hughes}}]{Benalcazar2017}%
  \BibitemOpen
  \bibfield  {author} {\bibinfo {author} {\bibfnamefont {W.~A.}\ \bibnamefont
  {Benalcazar}}, \bibinfo {author} {\bibfnamefont {B.~A.}\ \bibnamefont
  {Bernevig}},\ and\ \bibinfo {author} {\bibfnamefont {T.~L.}\ \bibnamefont
  {Hughes}},\ }\bibfield  {title} {\bibinfo {title} {Quantized electric
  multipole insulators},\ }\href {https://doi.org/10.1126/science.aah6442}
  {\bibfield  {journal} {\bibinfo  {journal} {Science}\ }\textbf {\bibinfo
  {volume} {357}},\ \bibinfo {pages} {61} (\bibinfo {year} {2017})}\BibitemShut
  {NoStop}%
\bibitem [{\citenamefont {Peterson}\ \emph {et~al.}(2018)\citenamefont
  {Peterson}, \citenamefont {Benalcazar}, \citenamefont {Hughes},\ and\
  \citenamefont {Bahl}}]{Peterson2018}%
  \BibitemOpen
  \bibfield  {author} {\bibinfo {author} {\bibfnamefont {C.~W.}\ \bibnamefont
  {Peterson}}, \bibinfo {author} {\bibfnamefont {W.~A.}\ \bibnamefont
  {Benalcazar}}, \bibinfo {author} {\bibfnamefont {T.~L.}\ \bibnamefont
  {Hughes}},\ and\ \bibinfo {author} {\bibfnamefont {G.}~\bibnamefont {Bahl}},\
  }\bibfield  {title} {\bibinfo {title} {A quantized microwave quadrupole
  insulator with topologically protected corner states},\ }\href
  {https://doi.org/10.1038/nature25777} {\bibfield  {journal} {\bibinfo
  {journal} {Nature}\ }\textbf {\bibinfo {volume} {555}},\ \bibinfo {pages}
  {346} (\bibinfo {year} {2018})}\BibitemShut {NoStop}%
\bibitem [{\citenamefont {{El Hassan}}\ \emph {et~al.}(2019)\citenamefont {{El
  Hassan}}, \citenamefont {Kunst}, \citenamefont {Moritz}, \citenamefont
  {Andler}, \citenamefont {Bergholtz},\ and\ \citenamefont
  {Bourennane}}]{ElHassan2019}%
  \BibitemOpen
  \bibfield  {author} {\bibinfo {author} {\bibfnamefont {A.}~\bibnamefont {{El
  Hassan}}}, \bibinfo {author} {\bibfnamefont {F.~K.}\ \bibnamefont {Kunst}},
  \bibinfo {author} {\bibfnamefont {A.}~\bibnamefont {Moritz}}, \bibinfo
  {author} {\bibfnamefont {G.}~\bibnamefont {Andler}}, \bibinfo {author}
  {\bibfnamefont {E.~J.}\ \bibnamefont {Bergholtz}},\ and\ \bibinfo {author}
  {\bibfnamefont {M.}~\bibnamefont {Bourennane}},\ }\bibfield  {title}
  {\bibinfo {title} {{Corner states of light in photonic waveguides}},\ }\href
  {https://doi.org/10.1038/s41566-019-0519-y} {\bibfield  {journal} {\bibinfo
  {journal} {Nat. Photonics}\ }\textbf {\bibinfo {volume} {13}},\ \bibinfo
  {pages} {697} (\bibinfo {year} {2019})}\BibitemShut {NoStop}%
\bibitem [{\citenamefont {Ge}(2017)}]{Ge2017}%
  \BibitemOpen
  \bibfield  {author} {\bibinfo {author} {\bibfnamefont {L.}~\bibnamefont
  {Ge}},\ }\bibfield  {title} {\bibinfo {title} {{Symmetry-protected zero-mode
  laser with a tunable spatial profile}},\ }\href
  {https://doi.org/10.1103/PhysRevA.95.023812} {\bibfield  {journal} {\bibinfo
  {journal} {Phys. Rev. A}\ }\textbf {\bibinfo {volume} {95}},\ \bibinfo
  {pages} {023812} (\bibinfo {year} {2017})}\BibitemShut {NoStop}%
\bibitem [{\citenamefont {Zhao}\ \emph {et~al.}(2018)\citenamefont {Zhao},
  \citenamefont {Miao}, \citenamefont {Teimourpour}, \citenamefont {Malzard},
  \citenamefont {El-Ganainy}, \citenamefont {Schomerus},\ and\ \citenamefont
  {Feng}}]{Zhao2018}%
  \BibitemOpen
  \bibfield  {author} {\bibinfo {author} {\bibfnamefont {H.}~\bibnamefont
  {Zhao}}, \bibinfo {author} {\bibfnamefont {P.}~\bibnamefont {Miao}}, \bibinfo
  {author} {\bibfnamefont {M.~H.}\ \bibnamefont {Teimourpour}}, \bibinfo
  {author} {\bibfnamefont {S.}~\bibnamefont {Malzard}}, \bibinfo {author}
  {\bibfnamefont {R.}~\bibnamefont {El-Ganainy}}, \bibinfo {author}
  {\bibfnamefont {H.}~\bibnamefont {Schomerus}},\ and\ \bibinfo {author}
  {\bibfnamefont {L.}~\bibnamefont {Feng}},\ }\bibfield  {title} {\bibinfo
  {title} {{Topological hybrid silicon microlasers}},\ }\href
  {https://doi.org/10.1038/s41467-018-03434-2} {\bibfield  {journal} {\bibinfo
  {journal} {Nat. Commun.}\ }\textbf {\bibinfo {volume} {9}},\ \bibinfo {pages}
  {981} (\bibinfo {year} {2018})}\BibitemShut {NoStop}%
\bibitem [{\citenamefont {Hodaei}\ \emph {et~al.}(2014)\citenamefont {Hodaei},
  \citenamefont {Miri}, \citenamefont {Heinrich}, \citenamefont
  {Christodoulides},\ and\ \citenamefont {Khajavikhan}}]{hodaei2014parity}%
  \BibitemOpen
  \bibfield  {author} {\bibinfo {author} {\bibfnamefont {H.}~\bibnamefont
  {Hodaei}}, \bibinfo {author} {\bibfnamefont {M.-A.}\ \bibnamefont {Miri}},
  \bibinfo {author} {\bibfnamefont {M.}~\bibnamefont {Heinrich}}, \bibinfo
  {author} {\bibfnamefont {D.~N.}\ \bibnamefont {Christodoulides}},\ and\
  \bibinfo {author} {\bibfnamefont {M.}~\bibnamefont {Khajavikhan}},\
  }\bibfield  {title} {\bibinfo {title} {Parity-time--symmetric microring
  lasers},\ }\href@noop {} {\bibfield  {journal} {\bibinfo  {journal}
  {Science}\ }\textbf {\bibinfo {volume} {346}},\ \bibinfo {pages} {975}
  (\bibinfo {year} {2014})}\BibitemShut {NoStop}%
\bibitem [{\citenamefont {Poli}\ \emph {et~al.}(2015)\citenamefont {Poli},
  \citenamefont {Bellec}, \citenamefont {Kuhl}, \citenamefont {Mortessagne},\
  and\ \citenamefont {Schomerus}}]{poli_selective_2015}%
  \BibitemOpen
  \bibfield  {author} {\bibinfo {author} {\bibfnamefont {C.}~\bibnamefont
  {Poli}}, \bibinfo {author} {\bibfnamefont {M.}~\bibnamefont {Bellec}},
  \bibinfo {author} {\bibfnamefont {U.}~\bibnamefont {Kuhl}}, \bibinfo {author}
  {\bibfnamefont {F.}~\bibnamefont {Mortessagne}},\ and\ \bibinfo {author}
  {\bibfnamefont {H.}~\bibnamefont {Schomerus}},\ }\bibfield  {title} {\bibinfo
  {title} {Selective enhancement of topologically induced interface states in a
  dielectric resonator chain},\ }\href {https://doi.org/10.1038/ncomms7710}
  {\bibfield  {journal} {\bibinfo  {journal} {Nat. Commun.}\ }\textbf {\bibinfo
  {volume} {6}},\ \bibinfo {pages} {7710} (\bibinfo {year} {2015})}\BibitemShut
  {NoStop}%
\bibitem [{\citenamefont {Peng}\ \emph {et~al.}(2014)\citenamefont {Peng},
  \citenamefont {Ozdemir}, \citenamefont {Lei}, \citenamefont {Monifi},
  \citenamefont {Gianfreda}, \citenamefont {Long}, \citenamefont {Fan},
  \citenamefont {Nori}, \citenamefont {Bender},\ and\ \citenamefont
  {Yang}}]{peng_parity-time-symmetric_2014}%
  \BibitemOpen
  \bibfield  {author} {\bibinfo {author} {\bibfnamefont {B.}~\bibnamefont
  {Peng}}, \bibinfo {author} {\bibfnamefont {S.~K.}\ \bibnamefont {Ozdemir}},
  \bibinfo {author} {\bibfnamefont {F.}~\bibnamefont {Lei}}, \bibinfo {author}
  {\bibfnamefont {F.}~\bibnamefont {Monifi}}, \bibinfo {author} {\bibfnamefont
  {M.}~\bibnamefont {Gianfreda}}, \bibinfo {author} {\bibfnamefont {G.~L.}\
  \bibnamefont {Long}}, \bibinfo {author} {\bibfnamefont {S.}~\bibnamefont
  {Fan}}, \bibinfo {author} {\bibfnamefont {F.}~\bibnamefont {Nori}}, \bibinfo
  {author} {\bibfnamefont {C.~M.}\ \bibnamefont {Bender}},\ and\ \bibinfo
  {author} {\bibfnamefont {L.}~\bibnamefont {Yang}},\ }\bibfield  {title}
  {\bibinfo {title} {Parity-time-symmetric whispering-gallery microcavities},\
  }\href {https://doi.org/10.1038/nphys2927} {\bibfield  {journal} {\bibinfo
  {journal} {Nat. Phys.}\ }\textbf {\bibinfo {volume} {10}},\ \bibinfo {pages}
  {394} (\bibinfo {year} {2014})}\BibitemShut {NoStop}%
\bibitem [{\citenamefont {Mertens}\ \emph {et~al.}(2005)\citenamefont
  {Mertens}, \citenamefont {Wehrspohn}, \citenamefont {Kitzerow}, \citenamefont
  {Matthias}, \citenamefont {Jamois},\ and\ \citenamefont
  {G{\"o}sele}}]{mertens2005tunable}%
  \BibitemOpen
  \bibfield  {author} {\bibinfo {author} {\bibfnamefont {G.}~\bibnamefont
  {Mertens}}, \bibinfo {author} {\bibfnamefont {R.}~\bibnamefont {Wehrspohn}},
  \bibinfo {author} {\bibfnamefont {H.-S.}\ \bibnamefont {Kitzerow}}, \bibinfo
  {author} {\bibfnamefont {S.}~\bibnamefont {Matthias}}, \bibinfo {author}
  {\bibfnamefont {C.}~\bibnamefont {Jamois}},\ and\ \bibinfo {author}
  {\bibfnamefont {U.}~\bibnamefont {G{\"o}sele}},\ }\bibfield  {title}
  {\bibinfo {title} {Tunable defect mode in a three-dimensional photonic
  crystal},\ }\href@noop {} {\bibfield  {journal} {\bibinfo  {journal} {Appl.
  Phys. Lett.}\ }\textbf {\bibinfo {volume} {87}},\ \bibinfo {pages} {241108}
  (\bibinfo {year} {2005})}\BibitemShut {NoStop}%
\bibitem [{\citenamefont {Kuzmiak}\ and\ \citenamefont
  {Maradudin}(1998)}]{kuzmiak1998localized}%
  \BibitemOpen
  \bibfield  {author} {\bibinfo {author} {\bibfnamefont {V.}~\bibnamefont
  {Kuzmiak}}\ and\ \bibinfo {author} {\bibfnamefont {A.~A.}\ \bibnamefont
  {Maradudin}},\ }\bibfield  {title} {\bibinfo {title} {Localized defect modes
  in a two-dimensional triangular photonic crystal},\ }\href@noop {} {\bibfield
   {journal} {\bibinfo  {journal} {Phys. Rev. B}\ }\textbf {\bibinfo {volume}
  {57}},\ \bibinfo {pages} {15242} (\bibinfo {year} {1998})}\BibitemShut
  {NoStop}%
\bibitem [{\citenamefont {Painter}\ \emph {et~al.}(1999)\citenamefont
  {Painter}, \citenamefont {Vu{\v{c}}kovi{\'c}},\ and\ \citenamefont
  {Scherer}}]{painter1999defect}%
  \BibitemOpen
  \bibfield  {author} {\bibinfo {author} {\bibfnamefont {O.}~\bibnamefont
  {Painter}}, \bibinfo {author} {\bibfnamefont {J.}~\bibnamefont
  {Vu{\v{c}}kovi{\'c}}},\ and\ \bibinfo {author} {\bibfnamefont
  {A.}~\bibnamefont {Scherer}},\ }\bibfield  {title} {\bibinfo {title} {Defect
  modes of a two-dimensional photonic crystal in an optically thin dielectric
  slab},\ }\href@noop {} {\bibfield  {journal} {\bibinfo  {journal} {J. Opt.
  Soc. Am. B}\ }\textbf {\bibinfo {volume} {16}},\ \bibinfo {pages} {275}
  (\bibinfo {year} {1999})}\BibitemShut {NoStop}%
\bibitem [{\citenamefont {Villeneuve}\ \emph {et~al.}(1996)\citenamefont
  {Villeneuve}, \citenamefont {Fan},\ and\ \citenamefont
  {Joannopoulos}}]{villeneuve1996microcavities}%
  \BibitemOpen
  \bibfield  {author} {\bibinfo {author} {\bibfnamefont {P.~R.}\ \bibnamefont
  {Villeneuve}}, \bibinfo {author} {\bibfnamefont {S.}~\bibnamefont {Fan}},\
  and\ \bibinfo {author} {\bibfnamefont {J.}~\bibnamefont {Joannopoulos}},\
  }\bibfield  {title} {\bibinfo {title} {Microcavities in photonic crystals:
  Mode symmetry, tunability, and coupling efficiency},\ }\href@noop {}
  {\bibfield  {journal} {\bibinfo  {journal} {Phys. Rev. B}\ }\textbf {\bibinfo
  {volume} {54}},\ \bibinfo {pages} {7837} (\bibinfo {year}
  {1996})}\BibitemShut {NoStop}%
\bibitem [{\citenamefont {Pan}\ \emph {et~al.}(2018)\citenamefont {Pan},
  \citenamefont {Zhao}, \citenamefont {Miao}, \citenamefont {Longhi},\ and\
  \citenamefont {Feng}}]{pan_photonic_2018}%
  \BibitemOpen
  \bibfield  {author} {\bibinfo {author} {\bibfnamefont {M.}~\bibnamefont
  {Pan}}, \bibinfo {author} {\bibfnamefont {H.}~\bibnamefont {Zhao}}, \bibinfo
  {author} {\bibfnamefont {P.}~\bibnamefont {Miao}}, \bibinfo {author}
  {\bibfnamefont {S.}~\bibnamefont {Longhi}},\ and\ \bibinfo {author}
  {\bibfnamefont {L.}~\bibnamefont {Feng}},\ }\bibfield  {title} {\bibinfo
  {title} {Photonic zero mode in a non-{Hermitian} photonic lattice},\ }\href
  {https://doi.org/10.1038/s41467-018-03822-8} {\bibfield  {journal} {\bibinfo
  {journal} {Nat. Commun.}\ }\textbf {\bibinfo {volume} {9}},\ \bibinfo {pages}
  {1308} (\bibinfo {year} {2018})}\BibitemShut {NoStop}%
\bibitem [{\citenamefont {Kittel}\ \emph {et~al.}(1996)\citenamefont {Kittel},
  \citenamefont {McEuen},\ and\ \citenamefont
  {McEuen}}]{kittel1996introduction}%
  \BibitemOpen
  \bibfield  {author} {\bibinfo {author} {\bibfnamefont {C.}~\bibnamefont
  {Kittel}}, \bibinfo {author} {\bibfnamefont {P.}~\bibnamefont {McEuen}},\
  and\ \bibinfo {author} {\bibfnamefont {P.}~\bibnamefont {McEuen}},\
  }\href@noop {} {\emph {\bibinfo {title} {Introduction to solid state
  physics}}},\ Vol.~\bibinfo {volume} {8}\ (\bibinfo  {publisher} {Wiley New
  York},\ \bibinfo {year} {1996})\BibitemShut {NoStop}%
\bibitem [{\citenamefont {Chaikin}\ \emph {et~al.}(1995)\citenamefont
  {Chaikin}, \citenamefont {Lubensky},\ and\ \citenamefont
  {Witten}}]{chaikin1995principles}%
  \BibitemOpen
  \bibfield  {author} {\bibinfo {author} {\bibfnamefont {P.~M.}\ \bibnamefont
  {Chaikin}}, \bibinfo {author} {\bibfnamefont {T.~C.}\ \bibnamefont
  {Lubensky}},\ and\ \bibinfo {author} {\bibfnamefont {T.~A.}\ \bibnamefont
  {Witten}},\ }\href@noop {} {\emph {\bibinfo {title} {Principles of condensed
  matter physics}}},\ Vol.~\bibinfo {volume} {10}\ (\bibinfo  {publisher}
  {Cambridge university press Cambridge},\ \bibinfo {year} {1995})\BibitemShut
  {NoStop}%
\bibitem [{\citenamefont {Feng}\ \emph {et~al.}(2017)\citenamefont {Feng},
  \citenamefont {El-Ganainy},\ and\ \citenamefont {Ge}}]{Feng2017}%
  \BibitemOpen
  \bibfield  {author} {\bibinfo {author} {\bibfnamefont {L.}~\bibnamefont
  {Feng}}, \bibinfo {author} {\bibfnamefont {R.}~\bibnamefont {El-Ganainy}},\
  and\ \bibinfo {author} {\bibfnamefont {L.}~\bibnamefont {Ge}},\ }\bibfield
  {title} {\bibinfo {title} {Non-{Hermitian} photonics based on parity–time
  symmetry},\ }\href {https://doi.org/10.1038/s41566-017-0031-1} {\bibfield
  {journal} {\bibinfo  {journal} {Nat. Photonics}\ }\textbf {\bibinfo {volume}
  {11}},\ \bibinfo {pages} {752} (\bibinfo {year} {2017})}\BibitemShut
  {NoStop}%
\bibitem [{\citenamefont {Rivero}\ and\ \citenamefont
  {Ge}(2021)}]{rivero2021chiral}%
  \BibitemOpen
  \bibfield  {author} {\bibinfo {author} {\bibfnamefont {J.~D.}\ \bibnamefont
  {Rivero}}\ and\ \bibinfo {author} {\bibfnamefont {L.}~\bibnamefont {Ge}},\
  }\bibfield  {title} {\bibinfo {title} {Chiral symmetry in non-hermitian
  systems: Product rule and clifford algebra},\ }\href@noop {} {\bibfield
  {journal} {\bibinfo  {journal} {Phys. Rev. B}\ }\textbf {\bibinfo {volume}
  {103}},\ \bibinfo {pages} {014111} (\bibinfo {year} {2021})}\BibitemShut
  {NoStop}%
\bibitem [{\citenamefont {Yoshida}\ \emph {et~al.}(2019)\citenamefont
  {Yoshida}, \citenamefont {Peters}, \citenamefont {Kawakami},\ and\
  \citenamefont {Hatsugai}}]{yoshida2019symmetry}%
  \BibitemOpen
  \bibfield  {author} {\bibinfo {author} {\bibfnamefont {T.}~\bibnamefont
  {Yoshida}}, \bibinfo {author} {\bibfnamefont {R.}~\bibnamefont {Peters}},
  \bibinfo {author} {\bibfnamefont {N.}~\bibnamefont {Kawakami}},\ and\
  \bibinfo {author} {\bibfnamefont {Y.}~\bibnamefont {Hatsugai}},\ }\bibfield
  {title} {\bibinfo {title} {Symmetry-protected exceptional rings in
  two-dimensional correlated systems with chiral symmetry},\ }\href@noop {}
  {\bibfield  {journal} {\bibinfo  {journal} {Phys. Rev. B}\ }\textbf {\bibinfo
  {volume} {99}},\ \bibinfo {pages} {121101} (\bibinfo {year}
  {2019})}\BibitemShut {NoStop}%
\bibitem [{\citenamefont {Lee}(2016)}]{lee2016anomalous}%
  \BibitemOpen
  \bibfield  {author} {\bibinfo {author} {\bibfnamefont {T.~E.}\ \bibnamefont
  {Lee}},\ }\bibfield  {title} {\bibinfo {title} {Anomalous edge state in a
  non-hermitian lattice},\ }\href@noop {} {\bibfield  {journal} {\bibinfo
  {journal} {Phys. Rev. Lett.}\ }\textbf {\bibinfo {volume} {116}},\ \bibinfo
  {pages} {133903} (\bibinfo {year} {2016})}\BibitemShut {NoStop}%
\bibitem [{\citenamefont {Yin}\ \emph {et~al.}(2018)\citenamefont {Yin},
  \citenamefont {Jiang}, \citenamefont {Li}, \citenamefont {L{\"u}},\ and\
  \citenamefont {Chen}}]{yin2018geometrical}%
  \BibitemOpen
  \bibfield  {author} {\bibinfo {author} {\bibfnamefont {C.}~\bibnamefont
  {Yin}}, \bibinfo {author} {\bibfnamefont {H.}~\bibnamefont {Jiang}}, \bibinfo
  {author} {\bibfnamefont {L.}~\bibnamefont {Li}}, \bibinfo {author}
  {\bibfnamefont {R.}~\bibnamefont {L{\"u}}},\ and\ \bibinfo {author}
  {\bibfnamefont {S.}~\bibnamefont {Chen}},\ }\bibfield  {title} {\bibinfo
  {title} {Geometrical meaning of winding number and its characterization of
  topological phases in one-dimensional chiral non-hermitian systems},\
  }\href@noop {} {\bibfield  {journal} {\bibinfo  {journal} {Phys. Rev. A}\
  }\textbf {\bibinfo {volume} {97}},\ \bibinfo {pages} {052115} (\bibinfo
  {year} {2018})}\BibitemShut {NoStop}%
\bibitem [{\citenamefont {Kawabata}\ \emph
  {et~al.}(2019{\natexlab{a}})\citenamefont {Kawabata}, \citenamefont
  {Shiozaki}, \citenamefont {Ueda},\ and\ \citenamefont {Sato}}]{Kawabata2019}%
  \BibitemOpen
  \bibfield  {author} {\bibinfo {author} {\bibfnamefont {K.}~\bibnamefont
  {Kawabata}}, \bibinfo {author} {\bibfnamefont {K.}~\bibnamefont {Shiozaki}},
  \bibinfo {author} {\bibfnamefont {M.}~\bibnamefont {Ueda}},\ and\ \bibinfo
  {author} {\bibfnamefont {M.}~\bibnamefont {Sato}},\ }\bibfield  {title}
  {\bibinfo {title} {{Symmetry and Topology in Non-Hermitian Physics}},\ }\href
  {https://doi.org/10.1103/PhysRevX.9.041015} {\bibfield  {journal} {\bibinfo
  {journal} {Phys. Rev. X}\ }\textbf {\bibinfo {volume} {9}},\ \bibinfo {pages}
  {41015} (\bibinfo {year} {2019}{\natexlab{a}})}\BibitemShut {NoStop}%
\bibitem [{\citenamefont {Rivero}\ and\ \citenamefont
  {Ge}(2020)}]{rivero_pseudochirality_2020}%
  \BibitemOpen
  \bibfield  {author} {\bibinfo {author} {\bibfnamefont {J.~D.}\ \bibnamefont
  {Rivero}}\ and\ \bibinfo {author} {\bibfnamefont {L.}~\bibnamefont {Ge}},\
  }\bibfield  {title} {\bibinfo {title} {Pseudochirality: {A} {Manifestation}
  of {Noether}'s {Theorem} in {Non}-{Hermitian} {Systems}},\ }\href
  {https://doi.org/10.1103/PhysRevLett.125.083902} {\bibfield  {journal}
  {\bibinfo  {journal} {Phys. Rev. Lett.}\ }\textbf {\bibinfo {volume} {125}},\
  \bibinfo {pages} {083902} (\bibinfo {year} {2020})}\BibitemShut {NoStop}%
\bibitem [{\citenamefont {Qi}\ \emph {et~al.}(2018)\citenamefont {Qi},
  \citenamefont {Zhang},\ and\ \citenamefont {Ge}}]{qi2018defect}%
  \BibitemOpen
  \bibfield  {author} {\bibinfo {author} {\bibfnamefont {B.}~\bibnamefont
  {Qi}}, \bibinfo {author} {\bibfnamefont {L.}~\bibnamefont {Zhang}},\ and\
  \bibinfo {author} {\bibfnamefont {L.}~\bibnamefont {Ge}},\ }\bibfield
  {title} {\bibinfo {title} {Defect states emerging from a non-hermitian
  flatband of photonic zero modes},\ }\href@noop {} {\bibfield  {journal}
  {\bibinfo  {journal} {Phys. Rev. Lett.}\ }\textbf {\bibinfo {volume} {120}},\
  \bibinfo {pages} {093901} (\bibinfo {year} {2018})}\BibitemShut {NoStop}%
\bibitem [{\citenamefont {Kawabata}\ \emph
  {et~al.}(2019{\natexlab{b}})\citenamefont {Kawabata}, \citenamefont
  {Higashikawa}, \citenamefont {Gong}, \citenamefont {Ashida},\ and\
  \citenamefont {Ueda}}]{kawabata2019topological}%
  \BibitemOpen
  \bibfield  {author} {\bibinfo {author} {\bibfnamefont {K.}~\bibnamefont
  {Kawabata}}, \bibinfo {author} {\bibfnamefont {S.}~\bibnamefont
  {Higashikawa}}, \bibinfo {author} {\bibfnamefont {Z.}~\bibnamefont {Gong}},
  \bibinfo {author} {\bibfnamefont {Y.}~\bibnamefont {Ashida}},\ and\ \bibinfo
  {author} {\bibfnamefont {M.}~\bibnamefont {Ueda}},\ }\bibfield  {title}
  {\bibinfo {title} {Topological unification of time-reversal and particle-hole
  symmetries in non-hermitian physics},\ }\href@noop {} {\bibfield  {journal}
  {\bibinfo  {journal} {Nat. Commun.}\ }\textbf {\bibinfo {volume} {10}},\
  \bibinfo {pages} {1} (\bibinfo {year} {2019}{\natexlab{b}})}\BibitemShut
  {NoStop}%
\bibitem [{\citenamefont {Okugawa}\ and\ \citenamefont
  {Yokoyama}(2019)}]{okugawa2019topological}%
  \BibitemOpen
  \bibfield  {author} {\bibinfo {author} {\bibfnamefont {R.}~\bibnamefont
  {Okugawa}}\ and\ \bibinfo {author} {\bibfnamefont {T.}~\bibnamefont
  {Yokoyama}},\ }\bibfield  {title} {\bibinfo {title} {Topological exceptional
  surfaces in non-hermitian systems with parity-time and parity-particle-hole
  symmetries},\ }\href@noop {} {\bibfield  {journal} {\bibinfo  {journal}
  {Phys. Rev. B}\ }\textbf {\bibinfo {volume} {99}},\ \bibinfo {pages} {041202}
  (\bibinfo {year} {2019})}\BibitemShut {NoStop}%
\bibitem [{\citenamefont {Wu}\ and\ \citenamefont
  {Hou}(2019)}]{wu2019symmetry}%
  \BibitemOpen
  \bibfield  {author} {\bibinfo {author} {\bibfnamefont {Y.-J.}\ \bibnamefont
  {Wu}}\ and\ \bibinfo {author} {\bibfnamefont {J.}~\bibnamefont {Hou}},\
  }\bibfield  {title} {\bibinfo {title} {Symmetry-protected localized states at
  defects in non-hermitian systems},\ }\href@noop {} {\bibfield  {journal}
  {\bibinfo  {journal} {Phys. Rev. A}\ }\textbf {\bibinfo {volume} {99}},\
  \bibinfo {pages} {062107} (\bibinfo {year} {2019})}\BibitemShut {NoStop}%
\bibitem [{\citenamefont {Ge}\ and\ \citenamefont
  {Türeci}(2013)}]{ge_antisymmetric_2013}%
  \BibitemOpen
  \bibfield  {author} {\bibinfo {author} {\bibfnamefont {L.}~\bibnamefont
  {Ge}}\ and\ \bibinfo {author} {\bibfnamefont {H.~E.}\ \bibnamefont
  {Türeci}},\ }\bibfield  {title} {\bibinfo {title} {Antisymmetric
  {PT}-photonic structures with balanced positive- and negative-index
  materials},\ }\href {https://doi.org/10.1103/PhysRevA.88.053810} {\bibfield
  {journal} {\bibinfo  {journal} {Phys. Rev. A}\ }\textbf {\bibinfo {volume}
  {88}},\ \bibinfo {pages} {053810} (\bibinfo {year} {2013})}\BibitemShut
  {NoStop}%
\bibitem [{\citenamefont {Zhang}\ \emph {et~al.}(2020)\citenamefont {Zhang},
  \citenamefont {Feng}, \citenamefont {Chen}, \citenamefont {Ge},\ and\
  \citenamefont {Wan}}]{zhang_synthetic_2020}%
  \BibitemOpen
  \bibfield  {author} {\bibinfo {author} {\bibfnamefont {F.}~\bibnamefont
  {Zhang}}, \bibinfo {author} {\bibfnamefont {Y.}~\bibnamefont {Feng}},
  \bibinfo {author} {\bibfnamefont {X.}~\bibnamefont {Chen}}, \bibinfo {author}
  {\bibfnamefont {L.}~\bibnamefont {Ge}},\ and\ \bibinfo {author}
  {\bibfnamefont {W.}~\bibnamefont {Wan}},\ }\bibfield  {title} {\bibinfo
  {title} {Synthetic {Anti}-{PT} {Symmetry} in a {Single} {Microcavity}},\
  }\href {https://doi.org/10.1103/PhysRevLett.124.053901} {\bibfield  {journal}
  {\bibinfo  {journal} {Phys. Rev. Lett.}\ }\textbf {\bibinfo {volume} {124}},\
  \bibinfo {pages} {053901} (\bibinfo {year} {2020})}\BibitemShut {NoStop}%
\bibitem [{\citenamefont
  {Scolarici}(2002)}]{scolarici_pseudoanti-hermitian_2002}%
  \BibitemOpen
  \bibfield  {author} {\bibinfo {author} {\bibfnamefont {G.}~\bibnamefont
  {Scolarici}},\ }\bibfield  {title} {\bibinfo {title} {Pseudoanti-{Hermitian}
  operators in quaternionic quantum mechanics},\ }\href
  {https://doi.org/10.1088/0305-4470/35/34/317} {\bibfield  {journal} {\bibinfo
   {journal} {J. Phys. A: Math. Gen.}\ }\textbf {\bibinfo {volume} {35}},\
  \bibinfo {pages} {7493} (\bibinfo {year} {2002})}\BibitemShut {NoStop}%
\bibitem [{\citenamefont {Chiu}\ \emph {et~al.}(2016)\citenamefont {Chiu},
  \citenamefont {Teo}, \citenamefont {Schnyder},\ and\ \citenamefont
  {Ryu}}]{Chiu2016}%
  \BibitemOpen
  \bibfield  {author} {\bibinfo {author} {\bibfnamefont {C.-K.}\ \bibnamefont
  {Chiu}}, \bibinfo {author} {\bibfnamefont {J.~C.}\ \bibnamefont {Teo}},
  \bibinfo {author} {\bibfnamefont {A.~P.}\ \bibnamefont {Schnyder}},\ and\
  \bibinfo {author} {\bibfnamefont {S.}~\bibnamefont {Ryu}},\ }\bibfield
  {title} {\bibinfo {title} {{Classification of topological quantum matter with
  symmetries}},\ }\href {https://doi.org/10.1103/RevModPhys.88.035005}
  {\bibfield  {journal} {\bibinfo  {journal} {Rev. Mod. Phys.}\ }\textbf
  {\bibinfo {volume} {88}},\ \bibinfo {pages} {035005} (\bibinfo {year}
  {2016})}\BibitemShut {NoStop}%
\bibitem [{\citenamefont {Malzard}\ \emph {et~al.}(2015)\citenamefont
  {Malzard}, \citenamefont {Poli},\ and\ \citenamefont
  {Schomerus}}]{Malzard2015}%
  \BibitemOpen
  \bibfield  {author} {\bibinfo {author} {\bibfnamefont {S.}~\bibnamefont
  {Malzard}}, \bibinfo {author} {\bibfnamefont {C.}~\bibnamefont {Poli}},\ and\
  \bibinfo {author} {\bibfnamefont {H.}~\bibnamefont {Schomerus}},\ }\bibfield
  {title} {\bibinfo {title} {{Topologically Protected Defect States in Open
  Photonic Systems with Non-Hermitian Charge-Conjugation and Parity-Time
  Symmetry}},\ }\href {https://doi.org/10.1103/PhysRevLett.115.200402}
  {\bibfield  {journal} {\bibinfo  {journal} {Phys. Rev. Lett.}\ }\textbf
  {\bibinfo {volume} {115}},\ \bibinfo {pages} {200402} (\bibinfo {year}
  {2015})}\BibitemShut {NoStop}%
\bibitem [{\citenamefont {Schomerus}(2013)}]{schomerus2013topologically}%
  \BibitemOpen
  \bibfield  {author} {\bibinfo {author} {\bibfnamefont {H.}~\bibnamefont
  {Schomerus}},\ }\bibfield  {title} {\bibinfo {title} {Topologically protected
  midgap states in complex photonic lattices},\ }\href@noop {} {\bibfield
  {journal} {\bibinfo  {journal} {Opt. Lett.}\ }\textbf {\bibinfo {volume}
  {38}},\ \bibinfo {pages} {1912} (\bibinfo {year} {2013})}\BibitemShut
  {NoStop}%
\bibitem [{\citenamefont {Weimann}\ \emph {et~al.}(2017)\citenamefont
  {Weimann}, \citenamefont {Kremer}, \citenamefont {Plotnik}, \citenamefont
  {Lumer}, \citenamefont {Nolte}, \citenamefont {Makris}, \citenamefont
  {Segev}, \citenamefont {Rechtsman},\ and\ \citenamefont
  {Szameit}}]{weimann2017topologically}%
  \BibitemOpen
  \bibfield  {author} {\bibinfo {author} {\bibfnamefont {S.}~\bibnamefont
  {Weimann}}, \bibinfo {author} {\bibfnamefont {M.}~\bibnamefont {Kremer}},
  \bibinfo {author} {\bibfnamefont {Y.}~\bibnamefont {Plotnik}}, \bibinfo
  {author} {\bibfnamefont {Y.}~\bibnamefont {Lumer}}, \bibinfo {author}
  {\bibfnamefont {S.}~\bibnamefont {Nolte}}, \bibinfo {author} {\bibfnamefont
  {K.~G.}\ \bibnamefont {Makris}}, \bibinfo {author} {\bibfnamefont
  {M.}~\bibnamefont {Segev}}, \bibinfo {author} {\bibfnamefont {M.~C.}\
  \bibnamefont {Rechtsman}},\ and\ \bibinfo {author} {\bibfnamefont
  {A.}~\bibnamefont {Szameit}},\ }\bibfield  {title} {\bibinfo {title}
  {Topologically protected bound states in photonic parity--time-symmetric
  crystals},\ }\href@noop {} {\bibfield  {journal} {\bibinfo  {journal} {Nat.
  Mater.}\ }\textbf {\bibinfo {volume} {16}},\ \bibinfo {pages} {433} (\bibinfo
  {year} {2017})}\BibitemShut {NoStop}%
\bibitem [{\citenamefont {Longhi}\ \emph {et~al.}(2015)\citenamefont {Longhi},
  \citenamefont {Gatti},\ and\ \citenamefont {Valle}}]{Longhi_robust_2015}%
  \BibitemOpen
  \bibfield  {author} {\bibinfo {author} {\bibfnamefont {S.}~\bibnamefont
  {Longhi}}, \bibinfo {author} {\bibfnamefont {D.}~\bibnamefont {Gatti}},\ and\
  \bibinfo {author} {\bibfnamefont {G.~D.}\ \bibnamefont {Valle}},\ }\bibfield
  {title} {\bibinfo {title} {Robust light transport in non-{Hermitian} photonic
  lattices},\ }\href {https://doi.org/10.1038/srep13376} {\bibfield  {journal}
  {\bibinfo  {journal} {Sci. Rep.}\ }\textbf {\bibinfo {volume} {5}},\ \bibinfo
  {pages} {13376} (\bibinfo {year} {2015})}\BibitemShut {NoStop}%
\bibitem [{\citenamefont {Zhang}\ \emph {et~al.}(2021)\citenamefont {Zhang},
  \citenamefont {Tian}, \citenamefont {Jiang}, \citenamefont {Lu},\ and\
  \citenamefont {Chen}}]{zhang_observation_2021}%
  \BibitemOpen
  \bibfield  {author} {\bibinfo {author} {\bibfnamefont {X.}~\bibnamefont
  {Zhang}}, \bibinfo {author} {\bibfnamefont {Y.}~\bibnamefont {Tian}},
  \bibinfo {author} {\bibfnamefont {J.-H.}\ \bibnamefont {Jiang}}, \bibinfo
  {author} {\bibfnamefont {M.-H.}\ \bibnamefont {Lu}},\ and\ \bibinfo {author}
  {\bibfnamefont {Y.-F.}\ \bibnamefont {Chen}},\ }\bibfield  {title} {\bibinfo
  {title} {Observation of higher-order non-{Hermitian} skin effect},\ }\href
  {https://doi.org/10.1038/s41467-021-25716-y} {\bibfield  {journal} {\bibinfo
  {journal} {Nat. Commun.}\ }\textbf {\bibinfo {volume} {12}},\ \bibinfo
  {pages} {5377} (\bibinfo {year} {2021})}\BibitemShut {NoStop}%
\bibitem [{\citenamefont {Wang}\ \emph {et~al.}(2022)\citenamefont {Wang},
  \citenamefont {Wang},\ and\ \citenamefont {Ma}}]{wang_non-hermitian_2022}%
  \BibitemOpen
  \bibfield  {author} {\bibinfo {author} {\bibfnamefont {W.}~\bibnamefont
  {Wang}}, \bibinfo {author} {\bibfnamefont {X.}~\bibnamefont {Wang}},\ and\
  \bibinfo {author} {\bibfnamefont {G.}~\bibnamefont {Ma}},\ }\bibfield
  {title} {\bibinfo {title} {Non-{Hermitian} morphing of topological modes},\
  }\href {https://doi.org/10.1038/s41586-022-04929-1} {\bibfield  {journal}
  {\bibinfo  {journal} {Nature}\ }\textbf {\bibinfo {volume} {608}},\ \bibinfo
  {pages} {50} (\bibinfo {year} {2022})}\BibitemShut {NoStop}%
\bibitem [{\citenamefont {Franca}\ \emph {et~al.}(2022)\citenamefont {Franca},
  \citenamefont {Könye}, \citenamefont {Hassler}, \citenamefont {van~den
  Brink},\ and\ \citenamefont {Fulga}}]{franca_non-hermitian_2022}%
  \BibitemOpen
  \bibfield  {author} {\bibinfo {author} {\bibfnamefont {S.}~\bibnamefont
  {Franca}}, \bibinfo {author} {\bibfnamefont {V.}~\bibnamefont {Könye}},
  \bibinfo {author} {\bibfnamefont {F.}~\bibnamefont {Hassler}}, \bibinfo
  {author} {\bibfnamefont {J.}~\bibnamefont {van~den Brink}},\ and\ \bibinfo
  {author} {\bibfnamefont {C.}~\bibnamefont {Fulga}},\ }\bibfield  {title}
  {\bibinfo {title} {Non-{Hermitian} {Physics} without {Gain} or {Loss}: {The}
  {Skin} {Effect} of {Reflected} {Waves}},\ }\href
  {https://doi.org/10.1103/PhysRevLett.129.086601} {\bibfield  {journal}
  {\bibinfo  {journal} {Phys. Rev. Lett.}\ }\textbf {\bibinfo {volume} {129}},\
  \bibinfo {pages} {086601} (\bibinfo {year} {2022})}\BibitemShut {NoStop}%
\bibitem [{\citenamefont {Gao}\ \emph {et~al.}(2023)\citenamefont {Gao},
  \citenamefont {Qiao}, \citenamefont {Pan}, \citenamefont {Wu}, \citenamefont
  {Yim}, \citenamefont {Chen}, \citenamefont {Midya}, \citenamefont {Ge},\ and\
  \citenamefont {Feng}}]{gao_two-dimensional_2023}%
  \BibitemOpen
  \bibfield  {author} {\bibinfo {author} {\bibfnamefont {Z.}~\bibnamefont
  {Gao}}, \bibinfo {author} {\bibfnamefont {X.}~\bibnamefont {Qiao}}, \bibinfo
  {author} {\bibfnamefont {M.}~\bibnamefont {Pan}}, \bibinfo {author}
  {\bibfnamefont {S.}~\bibnamefont {Wu}}, \bibinfo {author} {\bibfnamefont
  {J.}~\bibnamefont {Yim}}, \bibinfo {author} {\bibfnamefont {K.}~\bibnamefont
  {Chen}}, \bibinfo {author} {\bibfnamefont {B.}~\bibnamefont {Midya}},
  \bibinfo {author} {\bibfnamefont {L.}~\bibnamefont {Ge}},\ and\ \bibinfo
  {author} {\bibfnamefont {L.}~\bibnamefont {Feng}},\ }\bibfield  {title}
  {\bibinfo {title} {Two-{Dimensional} {Reconfigurable} {Non}-{Hermitian}
  {Gauged} {Laser} {Array}},\ }\href
  {https://doi.org/10.1103/PhysRevLett.130.263801} {\bibfield  {journal}
  {\bibinfo  {journal} {Phys. Rev. Lett.}\ }\textbf {\bibinfo {volume} {130}},\
  \bibinfo {pages} {263801} (\bibinfo {year} {2023})}\BibitemShut {NoStop}%
\bibitem [{\citenamefont {Hsu}\ \emph {et~al.}(2016)\citenamefont {Hsu},
  \citenamefont {Zhen}, \citenamefont {Stone}, \citenamefont {Joannopoulos},\
  and\ \citenamefont {Soljačić}}]{hsu_bound_2016}%
  \BibitemOpen
  \bibfield  {author} {\bibinfo {author} {\bibfnamefont {C.~W.}\ \bibnamefont
  {Hsu}}, \bibinfo {author} {\bibfnamefont {B.}~\bibnamefont {Zhen}}, \bibinfo
  {author} {\bibfnamefont {A.~D.}\ \bibnamefont {Stone}}, \bibinfo {author}
  {\bibfnamefont {J.~D.}\ \bibnamefont {Joannopoulos}},\ and\ \bibinfo {author}
  {\bibfnamefont {M.}~\bibnamefont {Soljačić}},\ }\bibfield  {title}
  {\bibinfo {title} {Bound states in the continuum},\ }\href
  {https://doi.org/10.1038/natrevmats.2016.48} {\bibfield  {journal} {\bibinfo
  {journal} {Nat. Rev. Mater.}\ }\textbf {\bibinfo {volume} {1}},\ \bibinfo
  {pages} {1} (\bibinfo {year} {2016})}\BibitemShut {NoStop}%
\bibitem [{\citenamefont {Longhi}(2014)}]{longhi_bound_2014}%
  \BibitemOpen
  \bibfield  {author} {\bibinfo {author} {\bibfnamefont {S.}~\bibnamefont
  {Longhi}},\ }\bibfield  {title} {\bibinfo {title} {Bound states in the
  continuum in {PT}-symmetric optical lattices},\ }\href
  {https://doi.org/10.1364/OL.39.001697} {\bibfield  {journal} {\bibinfo
  {journal} {Opt. Lett.}\ }\textbf {\bibinfo {volume} {39}},\ \bibinfo {pages}
  {1697} (\bibinfo {year} {2014})}\BibitemShut {NoStop}%
\bibitem [{\citenamefont {Bronson}(1991)}]{bronson1991matrix}%
  \BibitemOpen
  \bibfield  {author} {\bibinfo {author} {\bibfnamefont {R.}~\bibnamefont
  {Bronson}},\ }\href@noop {} {\emph {\bibinfo {title} {Matrix methods: An
  introduction}}}\ (\bibinfo  {publisher} {Gulf Professional Publishing},\
  \bibinfo {year} {1991})\BibitemShut {NoStop}%
\bibitem [{\citenamefont {Strang}(2016)}]{linearAlgebra}%
  \BibitemOpen
  \bibfield  {author} {\bibinfo {author} {\bibfnamefont {G.}~\bibnamefont
  {Strang}},\ }\href@noop {} {\emph {\bibinfo {title} {Introduction to {Linear}
  {Algebra}}}},\ \bibinfo {edition} {5th}\ ed.\ (\bibinfo  {publisher}
  {Wellesley-Cambridge Press},\ \bibinfo {address} {Wellesley},\ \bibinfo
  {year} {2016})\BibitemShut {NoStop}%
\bibitem [{SM()}]{SM}%
  \BibitemOpen
  \href@noop {} {}\bibinfo {note} {See the Supplemental Material for additional
  data and discussions, which includes Refs.~\cite{ge2023ssp} and
  \cite{qi2018linear}.}\BibitemShut {Stop}%
\bibitem [{\citenamefont {Miri}\ and\ \citenamefont
  {Alù}(2019)}]{miri_exceptional_2019}%
  \BibitemOpen
  \bibfield  {author} {\bibinfo {author} {\bibfnamefont {M.-A.}\ \bibnamefont
  {Miri}}\ and\ \bibinfo {author} {\bibfnamefont {A.}~\bibnamefont {Alù}},\
  }\bibfield  {title} {\bibinfo {title} {Exceptional points in optics and
  photonics},\ }\bibfield  {journal} {\bibinfo  {journal} {Science}\ }\textbf
  {\bibinfo {volume} {363}},\ \href {https://doi.org/10.1126/science.aar7709}
  {10.1126/science.aar7709} (\bibinfo {year} {2019})\BibitemShut {NoStop}%
\bibitem [{\citenamefont {Ezawa}(2018)}]{Kagome1}%
  \BibitemOpen
  \bibfield  {author} {\bibinfo {author} {\bibfnamefont {M.}~\bibnamefont
  {Ezawa}},\ }\bibfield  {title} {\bibinfo {title} {Higher-{Order}
  {Topological} {Insulators} and {Semimetals} on the {Breathing} {Kagome} and
  {Pyrochlore} {Lattices}},\ }\href
  {https://doi.org/10.1103/PhysRevLett.120.026801} {\bibfield  {journal}
  {\bibinfo  {journal} {Phys. Rev. Lett.}\ }\textbf {\bibinfo {volume} {120}},\
  \bibinfo {pages} {026801} (\bibinfo {year} {2018})}\BibitemShut {NoStop}%
\bibitem [{\citenamefont {Ni}\ \emph {et~al.}(2019)\citenamefont {Ni},
  \citenamefont {Weiner}, \citenamefont {Alù},\ and\ \citenamefont
  {Khanikaev}}]{Kagome2}%
  \BibitemOpen
  \bibfield  {author} {\bibinfo {author} {\bibfnamefont {X.}~\bibnamefont
  {Ni}}, \bibinfo {author} {\bibfnamefont {M.}~\bibnamefont {Weiner}}, \bibinfo
  {author} {\bibfnamefont {A.}~\bibnamefont {Alù}},\ and\ \bibinfo {author}
  {\bibfnamefont {A.~B.}\ \bibnamefont {Khanikaev}},\ }\bibfield  {title}
  {\bibinfo {title} {Observation of higher-order topological acoustic states
  protected by generalized chiral symmetry},\ }\href
  {https://doi.org/10.1038/s41563-018-0252-9} {\bibfield  {journal} {\bibinfo
  {journal} {Nat. Mater.}\ }\textbf {\bibinfo {volume} {18}},\ \bibinfo {pages}
  {113} (\bibinfo {year} {2019})}\BibitemShut {NoStop}%
\bibitem [{bib()}]{bibnote}%
  \BibitemOpen
  \href@noop {} {}\bibinfo {note} {One difference is that now the wave function
  at the top corner gains a phase factor of $\pi$.}\BibitemShut {Stop}%
\bibitem [{\citenamefont {Ge}\ \emph {et~al.}(2023)\citenamefont {Ge},
  \citenamefont {Gao},\ and\ \citenamefont {Feng}}]{ge2023ssp}%
  \BibitemOpen
  \bibfield  {author} {\bibinfo {author} {\bibfnamefont {L.}~\bibnamefont
  {Ge}}, \bibinfo {author} {\bibfnamefont {Z.}~\bibnamefont {Gao}},\ and\
  \bibinfo {author} {\bibfnamefont {L.}~\bibnamefont {Feng}},\ }\bibfield
  {title} {\bibinfo {title} {Non-{Hermitian} gauged laser arrays with localized
  excitations: {Anomalous} threshold and generalized principle of selective
  pumping},\ }\href {https://doi.org/10.1103/PhysRevB.108.104111} {\bibfield
  {journal} {\bibinfo  {journal} {Phys. Rev. B}\ }\textbf {\bibinfo {volume}
  {108}},\ \bibinfo {pages} {104111} (\bibinfo {year} {2023})}\BibitemShut
  {NoStop}%
\bibitem [{\citenamefont {Qi}\ and\ \citenamefont {Ge}(2023)}]{qi2018linear}%
  \BibitemOpen
  \bibfield  {author} {\bibinfo {author} {\bibfnamefont {B.}~\bibnamefont
  {Qi}}\ and\ \bibinfo {author} {\bibfnamefont {L.}~\bibnamefont {Ge}},\
  }\bibfield  {title} {\bibinfo {title} {Linear {Localization} of {Zero}
  {Modes} in {Weakly} {Coupled} {Non}-{Hermitian} {Reservoirs}},\ }\href
  {https://doi.org/10.1002/apxr.202300066} {\bibfield  {journal} {\bibinfo
  {journal} {Adv. Phys. Res.}\ }\textbf {\bibinfo {volume} {2023}},\ \bibinfo
  {pages} {2300066} (\bibinfo {year} {2023})}\BibitemShut {NoStop}%
\end{thebibliography}%

\end{document}